\documentclass[12pt]{article}

\usepackage{graphicx}

\newcounter{lastnote}

\title{Magneto-orbital helices and multiferroicity} 

\author{\normalsize{N. J. Perks ,$^{1\ast}$ R. D. Johnson,$^{1,2}$ C. Martin,$^{3}$ L. C. Chapon,$^{4}$ P. G Radaelli$^{1}$}\\
\\
\scriptsize{$^{1}$Clarendon Laboratory, Department of Physics, University of Oxford, Oxford, OX1 3PU, UK}\\
\scriptsize{$^{2}$ISIS facility, Rutherford Appleton Laboratory-STFC, Chilton, Didcot, OX11 0QX, UK}\\
\scriptsize{$^{3}$Laboratoire CRISMAT, ENSICAEN, UMR F-6508 CNRS, 6 Boulevard du Marechal Juin, F-14050 Caen, France}\\
\scriptsize{$^{4}$Institut Laue-Langevin, BP 156X, 38042 Grenoble, France}\\
\\
\small{$^\ast$To whom correspondence should be addressed; E-mail: natasha.perks@physics.ox.ac.uk.}
}

\date{}

\begin{document} 

\baselineskip24pt

\maketitle

\newpage

\textbf{Orbitally ordered states evolving coincidentally with structural distortions and magnetic ordering provide a novel route to controlling electronic and magnetic properties of materials through external fields.  We report an unprecedented magneto-orbital helix in CaMn$_7$O$_{12}$, found to give rise to the largest magnetically induced ferroelectric polarisation measured to date.  Characterisation of the structural modulation using x-ray diffraction, and analysis of magnetic exchange shows that orbital order is crucial in stabilising a chiral magnetic structure, thus allowing for electric polarisation.  Additionally, the presence of a global structural rotation enables the coupling between this polarisation and magnetic helicity required for multiferrocitiy.  These novel principles open up the possibility of discovering new high-temperature multiferroics.}
\\

Orbital ordering phenomena provide one of the most striking manifestations of electronic correlations in a variety of materials, most notably in 3d transition metal oxides \cite{hotta2006}.  In these systems, below a certain temperature T$_{OO}$, a distinct anisotropic pattern of orbital occupation emerges from a nearly isotropic high-temperature state.  Since this state is strongly connected with both structural distortions (the cooperative Jahn Teller effect) and with magnetism, orbital ordering has been regarded for many years as a promising ingredient to allow the cross-coupling of electric properties.  For example, colossal magnetoresistance takes place because an orbitally ordered, insulating and antiferromagnetic state competes with a ferromagnetic isotropic metallic state, and its onset can be controlled by the application of an external magnetic field \cite{ramirez1997}.  Recently, it has been proposed that some charge and orbitally ordered states could be electrically polar, so that orbital ordering could promote high-temperature, magnetically controllable ferroelectricity \cite{khomskii2008}.  Here, we show that a unique form of incommensurate orbital ordering in the complex manganese oxide CaMn$_7$O$_{12}$, appearing below 250 K, is responsible for the largest observed magnetically induced ferroelectric polarisation.  Unlike all other known systems, orbital occupation in CaMn$_7$O$_{12}$ does not select fixed preferential directions, but  rotates around the 3-fold crystallographic axis, preserving the overall point-group symmetry.  The resulting pattern of magnetic exchange interactions produces a magnetic helix having twice the period of the orbital modulation below 90 K.  This magnetic texture breaks inversion symmetry and couples to the crystal structure via the relativistic antisymmetric exchange to yield a large electrical polarisation.

Figure 1a shows the room temperature crystal structure of CaMn$_7$O$_{12}$.  It can be described as a perovskite superstructure, with 3/4 of the $A$ sites occupied by Mn$^{3+}$ (hereafter labelled Mn1), and the remaining occupied by Ca.  A trigonal structure (space group $R\bar{3}$) develops on cooling below $\sim$ 400 K, and is best described by considering the pseudo-cubic B sites, which form three inequivalent triangular layers stacked along the hexagonal $c$ axis (shown in Figure 1a, and coloured green, blue and red throughout this report).  Within each layer, 1/4 of the B sites retain $\bar{3}$ symmetry (Mn3, black-outlined circles in Figure 1a), whereas the other 3/4 have $\bar{1}$ symmetry (Mn2).  Charge ordering develops gradually on further cooling, with Mn2 and Mn3 tending towards Mn$^{3+}$ and Mn$^{4+}$, respectively.  An incommensurate structural modulation along the hexagonal \emph{c} axis appears below T$_{OO}$=250K, with a propagation vector \emph{q}$_{c}$  = (0, 0, 2.077$\pm0.001$) at 150 K, producing a continuous variation in Mn-O bond lengths \cite{slawinski2009}.

The magnetic structure below ${T_\mathrm{N}}$ = 90 K was determined using neutron powder diffraction \cite{johnson2012}, and is shown in Figure 1b.  Magnetic anisotropy results in the spins lying within the hexagonal \emph{ab} plane.  Within each of the triangular layers, Mn2 spins are parallel to each other, whilst Mn3 spins are approximately antiparallel.  Adjacent layers are rotated by 124 $^\circ$ in the ab plane with respect to each other, leading to an incommensurate magnetic modulation with propagation vector \emph{q}$_{m}$ = (0, 0, 1.037) (Supplementary Section SI).  As observed in many magnetically ordered crystals (see, for example \cite{noda2008}), below ${T_\mathrm{N}}$,  \emph{q}$_{c}$ = 2\emph{q}$_{m}$.  In the present case, the structural modulation is also present in the paramagnetic phase, and the magnetic modulation "locks in" to it as it develops at ${T_\mathrm{N}}$.

We now focus on the incommensurate structural modulation appearing below T$_{OO}$, and demonstrate that this corresponds to the onset of a novel form of orbital ordering.  We performed Rietveld refinement using \emph{JANA}2006 \cite{petricek2006}, of single crystal diffraction data measured at 150 K (Supplementary Section SII).  Consistent with previous reports \cite{slawinski2009}, we found that the modulated crystal structure is described by the centrosymmetric four-dimensional space group $R\bar{3}$(00$\gamma$)0 with propagation vector \emph{q}$_{c}$ = (0, 0, 2.077).  In Fig 2a, we plot the Mn-O bond lengths for adjacent unit cells along the $c$ axis around Mn2 (the variation in the Mn1-O and Mn3-O bond lengths are small and dominated by movement of the oxygen anions (Supplementary Section SIII)).  This structural modulation has strong implications for the nature of orbital occupation as follows.  We define orthogonal $x$, $y$ and $z$ bond directions, schematically shown in Figure 1c.  In the unmodulated, average structure, the Mn$^{3+}$O$_6$ octahedra around Mn2 are compressed, giving two short and four long bonds, implying an unusual occupation of the ${x^{2}-y^{2}}$ orbital.  In reality, the structural modulation leads to an incommensurate rotation of a single pair of long bonds in the \emph{xy} plane, consistent with a periodic occupation of the ${3x^{2}-r^2}$ and ${3y^{2}-r^2}$ orbitals, shown at the side of Figure 2.  We can further characterise this orbital rotation by employing a formalism \cite{goodenough2001} \cite{wu2011}, first proposed by Goodenough, which quantifies orbital occupation according to a mixing angle, tan$\theta$ = $\sqrt{3}$${(x-z)/(2y-x-z)}$.  Following the conventions in \cite{wu2011}, we need only to consider the angular sector $2\pi/3 \leq \theta \leq 4\pi/3$ (inset to Figure 2).  As shown in Fig 2b, the orbital occupation rotates periodically between the \emph{${3x^{2}-r^2}$} and \emph{${3y^{2}-r^2}$} orbitals, passing through the  ${x^{2}-y^{2}}$  only at the nodal points.   Figure 2c shows the Mn2 charge modulation accompanying the orbital modulation, calculated using the Bond Valence Sum method \cite{brown1985}.  Valence fluctuations are surprisingly small (~1 $\%$ --- Supplementary Section SIV), indicating that the modulation affects predominantly the orbital occupation.  

Although orbital ordering is a widespread phenomenon in manganese oxides \cite{radaelli1997}\cite{feiner1999}, an incommensurate modulation where the orbital occupation rotates \emph{perpendicular} to the propagation direction is unprecedented.  The resulting orbital arrangement has a profound effect on magnetic ordering below T$_N=90 K$.  This can be seen by applying the Goodenough-Kanamori-Anderson (GKA) rules \cite{anderson1950} \cite{kanamori1957_1} \cite{kanamori1957_2} \cite{goodenough1958}, which predict the nature of exchange interactions.  In the absence of the structural modulation, the pseudo-perovskite structure is heavily distorted, and antiferromagnetic interactions are weak due to the large departure of the Mn-O-Mn bond angles from 180$^{\circ}$.  The alternation of filled and empty ${x^{2}-y^{2}}$ orbitals on Mn2 and Mn3 promotes ferromagnetic exchange, and a net ${ferromagnetic}$ structure is predicted.  The situation changes drastically in the presence of the orbital modulation, as the magnetic interaction between adjacent A, B and C layers is predominantly mediated by the orbitals that are most affected by the modulation.  This is shown schematically in Fig 3a, which depicts a region of the crystal structure where the modulation amplitude is largest.  Mn3 in the B layer is connected to the Mn2 sites in the A layer by filled ${3x^{2}-r^2}$ orbitals, yielding a \emph{ferromagnetic} interaction $J_1$, and to the Mn2 sites in the C layer by empty orbitals, yielding an \emph{antiferromagnetic} interaction $J_2$.  This produces an effective \emph{ferromagnetic} interaction between spins in each of the A, B and C layers, as all $J_1$($J_2$)'s are equivalent by symmetry (shown in Figure 3b).  Across layers there is an effective next-nearest neighbour \emph{antiferromagnetic} interaction $J_{12}$, which is strongest at the "crest" of the modulation, but remains antiferromagnetic everywhere except possibly at the nodal point.  The overall nearest-neighbour interaction between layers, $J_3$, results from the competition between full-empty ($x$-$z$) orbitals and empty-empty ($y$-$z$) orbitals, which are ferro- and antiferromagnetic, respectively.  Although the sign of $J_3$ cannot be predicted \emph{a priori}, it is experimentally found to be positive (antiferromagnetic).  The magnetic helix therefore results from the competition between antiferromagnetic nearest-neighbour and next-nearest neighbour interactions.  We can test this hypothesis with a very simple model, which captures the effect of the average interactions $J_1$, $J_2$ and $J_3$, assuming a magnetic propagation vector (0,0,q$_z$) and identical spins $S$ on Mn2 and Mn3 sites.  The magnetic Heisenberg exchange energy per Mn3 site can be written as:

\begin{equation}
\label{eq: energy}
E=S^2\left[3J_1 \cos \phi +3 J_2 \cos (\phi-2\psi) +6 J_3 \cos \psi\right]
\end{equation}

where $\psi=2/3 \pi q_z$ and $\phi$ are the angles between Mn2-Mn2 and Mn3-Mn2 spins in adjacent layers, respectively.  Minimising the energy in eq. \ref{eq: energy} yields the phase diagram shown in Figure 3c.  The wide incommensurate phase field is bound on either side by two distinct commensurate collinear phases, both with $q_z$ = 1.5.  The white line indicates the range of parameters for which the most stable propagation vector is that determined experimentally.  The red dot corresponds precisely to the experimental magnetic structure ($\phi = 31 ^{\circ}$), and is found for J$_2$/J$_1$= -0.85 and J$_3$/J$_1$=-0.63, in good agreement with our qualitative considerations based on the GKA rules.  The coupling between Mn1 and adjacent Mn2 sites is expected (and observed) to be ferromagnetic (Figure 3a) since the sites are linked via oxygen anions with highly-distorted bonds at less than the GKA critical angle.  The most important ingredient not included in our simple model is the  \emph{q}$_{c}$ = 2\emph{q}$_{m}$ lock-in term, which intertwines the magnetic and orbital modulation to form a \emph{magneto-orbital helix}.  Phenomenologically, different forms of the lock-in term are allowed by symmetry, with the most likely candidate involving a third-harmonic "squaring up" of the magnetic modulation.

Below T$_N$, CaMn$_7$O$_{12}$ acquires the largest measured magnetically-induced electric polarisation along the $c$ axis \cite{johnson2012}, which can be explained by the existence of the magneto-orbital helix.  As CaMn$_7$O$_{12}$ has ferroaxial point symmetry $\bar{3}$, the electrical polarisation can arise through coupling between the magneto-orbital helix and a global rotation between different elements of the crystal structure, as shown in Figure 4a:  the 6 oxygen atoms surrounding an individual Mn3 are all rotated in the same clockwise direction.  This is described by the phenomenological invariant $\alpha\sigma \bf{A} \cdot \bf{P}$, where $\alpha$ is a coupling constant,  $\sigma$ is the magnetic helicity, $\bf P$ is the electrical polarisation and $\bf A$ is an axial vector representing the structural rotation.  The magnetic helicity is not invariant by a global rotation of the spins, so the coupling  can only be of relativistic nature.  The relevant mechanism is the antisymmetric Dzyaloshinskii-Moriya (DM) exchange, where energy can be gained by distorting the crystal structure and/or the electronic density in the presence of non-collinear spin configurations.  Although both structural and electronic distortions are allowed in the present case, we will employ the former to illustrate how this mechanism can produce electrical polarisation in CaMn$_7$O$_{12}$.  When two metal sites carrying non-collinear spins are joined by common ligand atoms (in this case Mn--O--Mn), energy can be gained  by displacing the ligand through a vector $\bf u$ so that $\Delta E = \lambda \bf{u} \cdot [\bf{r}_{12} \times (\bf{S}_1 \times \bf{S}_2)]=\lambda \bf{D} \cdot (\bf{S}_1 \times \bf{S}_2) $, where $\bf{S}_1$ and $\bf{S}_2$ are the spins on the two sites, $\bf{r}_{12}$ is the position vector connecting them and $\lambda$ is a coupling constant.  $\bf{D}=\bf{u} \times \bf{r}_{12}$  is the familiar DM vector \cite{dzyaloshinskii1964}\cite{moriya1960}.  Energy is therefore minimised by a pattern of local ligand (oxygen) displacements $\bf u$ associated with pairs of spins and either parallel or antiparallel to the vectors $\bf{r}_{12} \times (\bf{S}_1 \times \bf{S}_2)$, depending on the sign of $\lambda$.  Fig 4b shows the pattern of $\bf{r}_{12} \times (\bf{S}_1 \times \bf{S}_2)$ vectors (black arrows) for clusters of Mn2 around a single Mn3.  Assuming $\lambda<0$, the DM interaction will favour oxygen displacements \emph{parallel} to these arrows, making the Mn$^{3+}$-O-Mn$^{4+}$ bonds between the red and the blue layer \emph{flatter} (i.e., the bond angles closer to 180$^{\circ}$) and Mn$^{3+}$-O-Mn$^{4+}$ bonds between the red and the green layer \emph{more acute} (i.e., the bond angles farther from 180$^{\circ}$).  It is easy to see that this pattern of displacements will result in the central Mn$^{4+}$ moving \emph{upwards} (yellow arrow), creating a local polarisation along the positive $c$ direction (identical considerations for $\lambda >0$ lead to a polarisation along the negative $c$ direction).  Since the magnetic structure is globally chiral, these local polarisations add up to give a net electrical polarisation.  It is important to stress that such a simple relation between local $\bf u$ and global $\bf P$ vectors is only possible in the presence of a global structural rotation, which is allowed in point group $\bar{3}$, as no net polarisation would arise for 180$^{\circ}$ Mn$^{3+}$-O-Mn$^{4+}$ bonds. Ferroaxiality is thus an essential ingredient for the development of electrical polarisation in CaMn$_7$O$_{12}$.  The large value of $P$ is readily explained by considering the angle of rotation $\phi$ between the spins  $\bf{S}_1$ and $\bf{S}_2$,  since $P$ should be roughly proportional to $\sin \phi$.  The average angle of rotation is $\langle \phi \rangle=124/2^{\circ}=62^{\circ}$ whereas for the prototypical magnetic multiferroic TbMnO$_3$, $\phi=200^{\circ}$, so that $\sin \langle \phi \rangle$ in CaMn$_7$O$_{12}$ is 2.5 times larger than in TbMnO$_3$, in relatively good agreement with that found experimentally ($P$ four times larger), considering the differences in geometry and coupling.  The small deviation from commensurability is \emph{not} an important factor in determining the electrical properties of CaMn$_7$O$_{12}$:   $P$ would be almost as large if the magnetic propagation vector was  exactly (0,0,1), since what matters is the large rotation angle between spins in adjacent A, B and C layers.

To conclude, we have shown that the magnetic and structural modulations in CaMn$_7$O$_{12}$ are intertwined to form an  \emph{incommensurate magneto-orbital helix} .  This unprecedented texture represents a novel manifestation of the deep connection between orbital physics and magnetism in transition-metal oxides.  The incommensurate structural modulation is crucial in stabilising a chiral magnetic structure, which breaks inversion symmetry, producing the largest magnetically-induced electrical polarisation ever observed in any system.  We have also shown how the presence of a global structural rotation (ferroaxiality) is essential in promoting the coupling between magneto-orbital helicity and ferroelectricity.  These observations open up the possibility of finding a new class of multiferroic materials in which giant ferroelectric polarisations may arise, underpinning the development of multiferroic technology.      


The work done at the University of Oxford was funded by an EPSRC grant number EP/J003557/1, entitled "New Concepts in Multiferroics and Magnetoelectrics".  
\\
\setlength{\parindent}{0in}

\newpage
\begin{figure}
\begin{center}
\includegraphics[width=12cm]{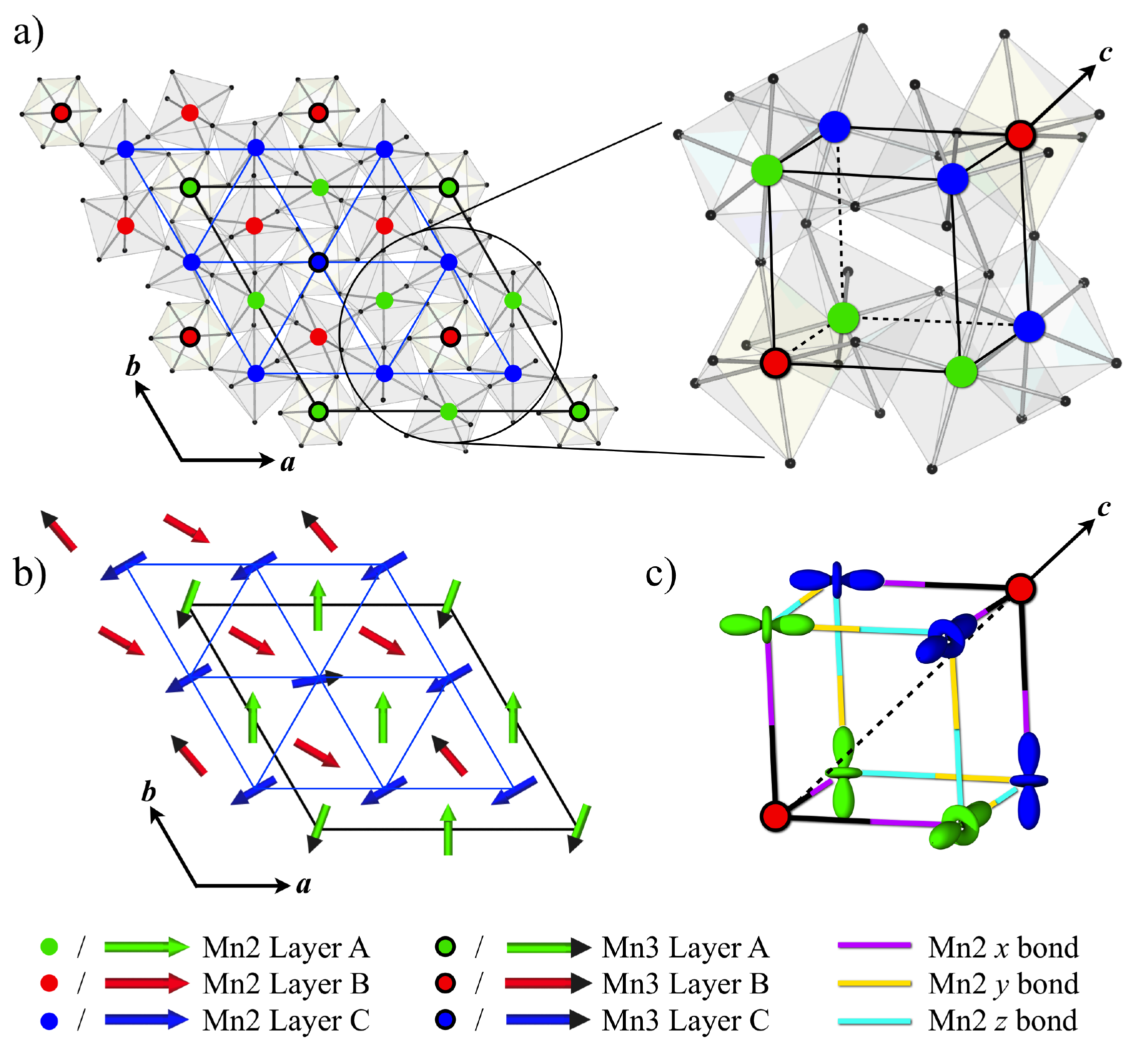}
\end{center}
\end{figure}
\textbf{Figure 1  The crystal structure, magnetic structure and associated orbital orientations of CaMn$_7$O$_{12}$.}  \textbf{(a)} (left panel) The room temperature crystal structure in the hexagonal setting, highlighting the three triangular layers (A=blue, B=red and C=green) stacked along the hexagonal $c$ axis (cubic [111] direction).  The structure can also be described within the pseudocubic perovskite setting (right panel), with Mn3 sites (black-outlines spheres, with nominal valence 4+) occupying the apex sites along the [111] diagonal.  \textbf{(b)} The magnetic structure at 90 K.  The triangular layers are again shown as blue, red and green, with a 124$^{\circ}$ spin rotation between adjacent layers.  \textbf{(c)} A schematic showing the B sites in the pseudocubic setting, defining the $x$, $y$ and $z$ bond directions which are used to describe orbital orientation.     \\ \\

\begin{figure}
\begin{center}
\includegraphics[width=8.5cm]{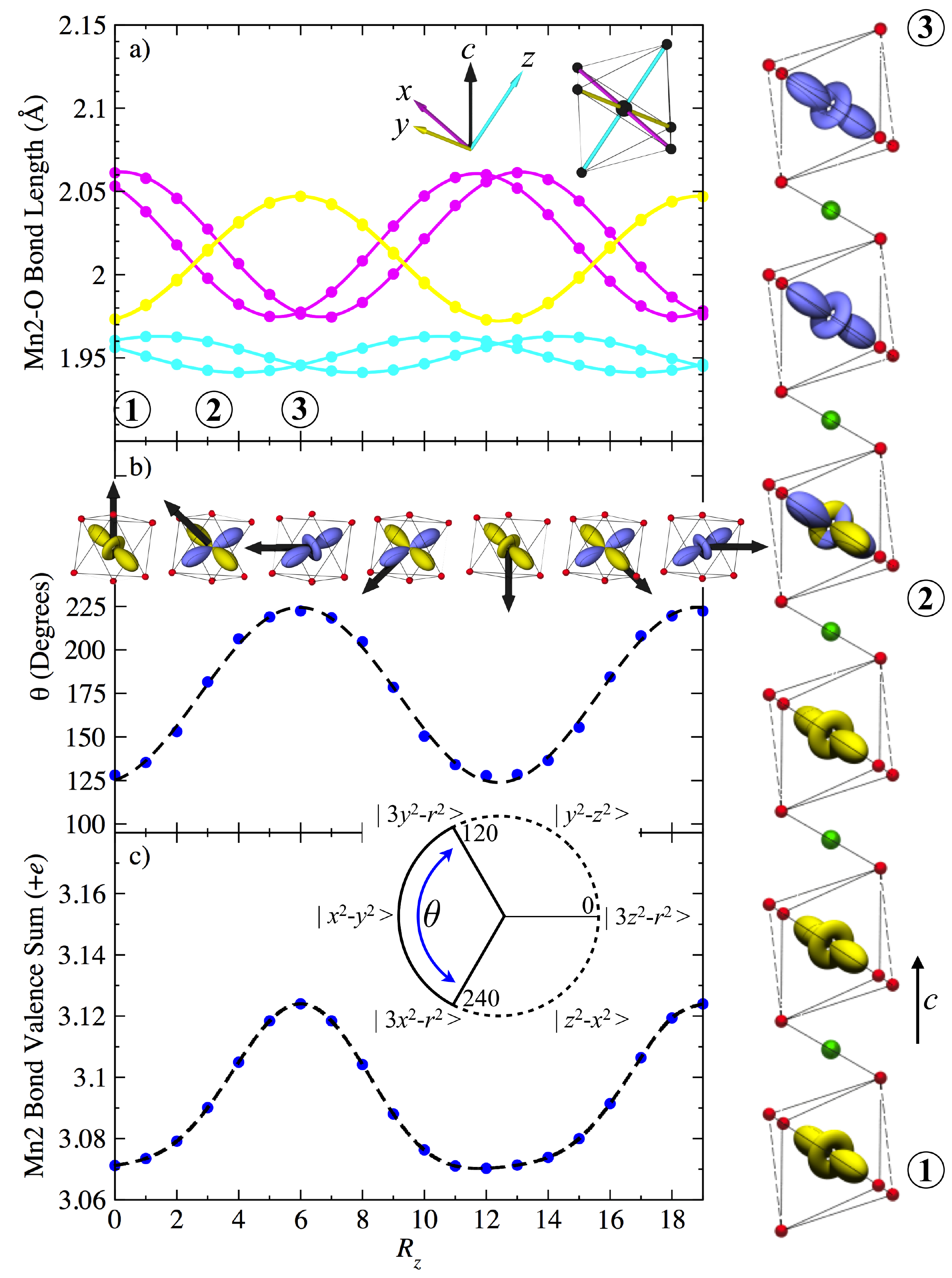}
\end{center}
\end{figure}
\textbf{Figure 2 The incommensurate structural modulation coupled to orbital rotation and valence fluctuations}  \textbf{(a)}  Variation in Mn2-O bond lengths along $x$, $y$ and $z$ (coloured according to the schematic in Figure 1c) as a function of $R_z$, plotted across 19 unit cells along the $c$ axis.  \textbf{(b)} Orbital mixing angle $\theta$ for the Mn2 octahedra as a function of $R_z$.  $\theta$ correlates directly with orbital occupation, as shown by the circular inset.  The orbital rotation is shown, along with the coupled helical spin rotation.  \textbf{(c)}  The Mn2 charge modulation, which accompanies the structural modulation along the $c$ axis.  The incommensurate orbital modulation for half a period of the structural modulation is depicted at the side of the figure, with positions 1,2 and 3 corresponding to the labelled points in (a) . \\ \\              

\begin{figure}
\begin{center}
\includegraphics[width=12cm]{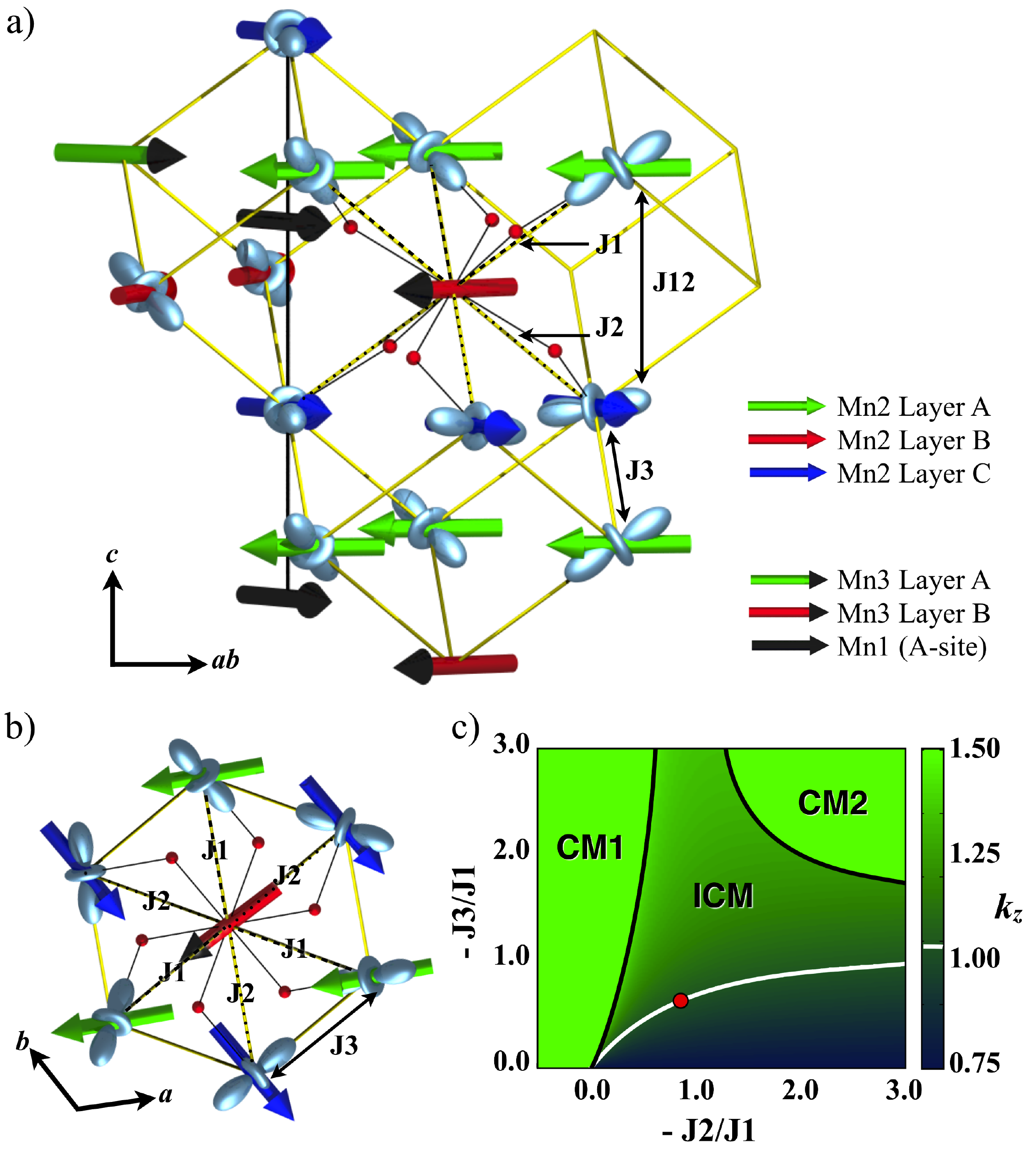}
\end{center}
\end{figure}
\textbf{Figure 3 The effect of orbital ordering on magnetic exchange interactions, promoting a chiral magnetic structure}  \textbf{(a)} Three pseudocubic unit cells illustrating the coupling between the incommensurate orbital modulation and the magnetic structure.  The occupied Mn2 (Mn$^{3+}$) orbitals are shown in grey, giving rise to ferromagnetic exchange interactions ($J_1$), shown by a dashed line, and antiferromagnetic exchange interactions ($J_2$), shown by a dotted line.  The nearest-neighbour and effective next-nearest neighbour antiferromagnetic exchanges between layers are labelled $J_{3}$ and $J_{12}$, respectively.  The ferromagnetic exchange along the Mn1---Mn2 chains is shown by a solid black line.  \textbf{(b)} Figure 3a viewed down the \emph{c} axis, showing the symmetry equivalent exchange interactions.  \textbf{(c)}  Phase diagram of the CaMn$_7$O$_{12}$ magnetic structures, obtained by minimising the classical Heisenberg exchange energy in eq. 1 as a function of the angles $\psi$ and $\phi$ for different exchange constants J$_1$, $J_2$ and J$_3$.  Spins on Mn2 and Mn3 sites were taken to be equal (different spins would be equivalent to renormalising the exchange constants).  CM1 and CM2 are two commensurate collinear phases, both with propagation vector (0, 0, 1.5), in which the Mn2 and Mn3 in each layer are antiparallel (for CM1) and parallel (for CM2).  ICM is an incommensurate phase with rotating spins (magnetic helix).  The shading indicates the value of the $z$-component of the propagation vector.  The white line indicates the locus of points with the experimental propagation vector (0, 0, 1.037).  The red dot corresponds to the experimental magnetic structure ($\phi$=31$^{\circ}$). \\ \\

\newpage
\begin{figure}
\begin{center}
\includegraphics[width=9cm]{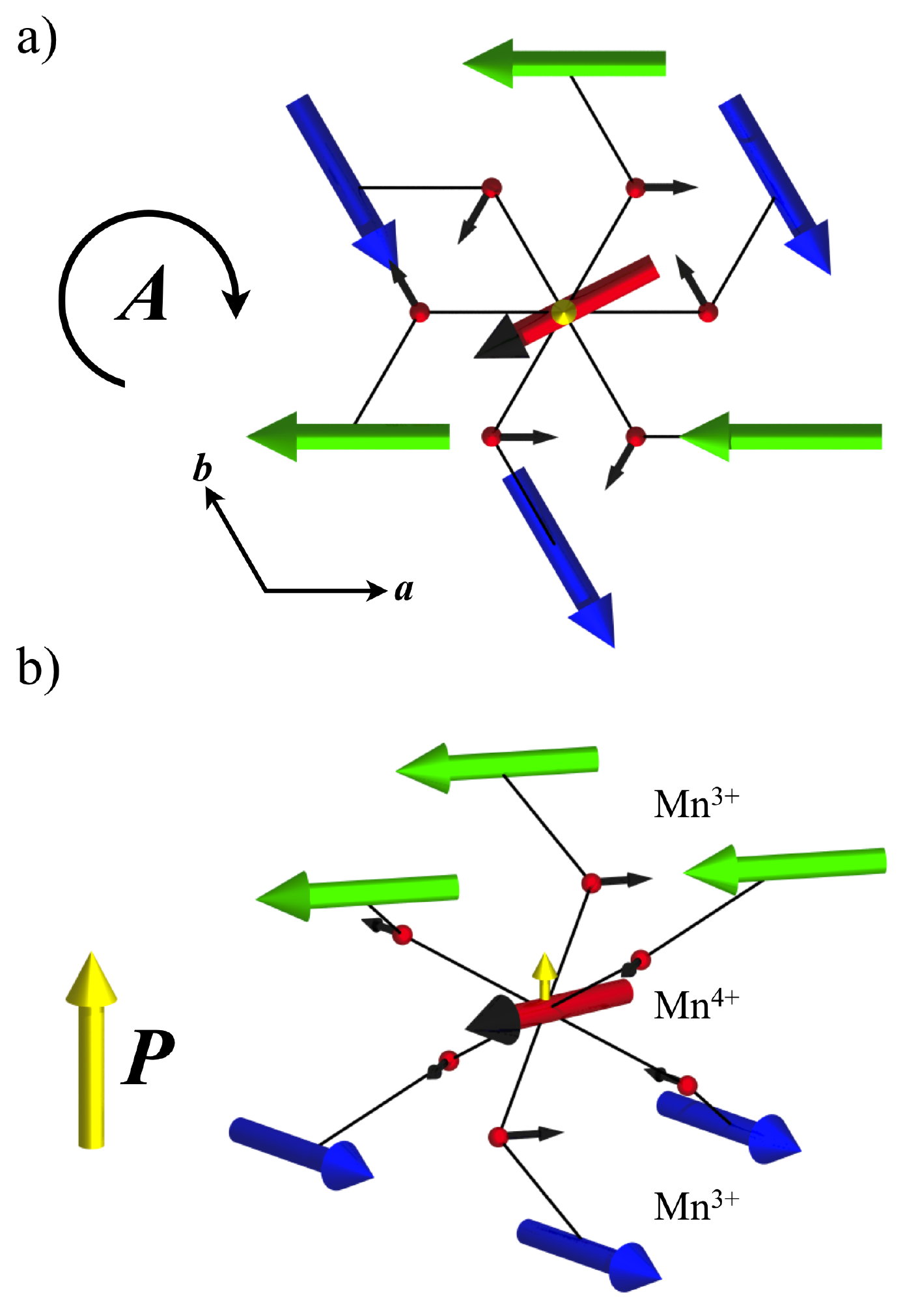}
\end{center}
\end{figure}
\textbf{Figure 4 The coupling between magnetic helicity and ferroelectricity, leading to a giant polarisation}.  \textbf{(a)}  The global structural rotation described by the axial vector $A$ along the $c$ axis is evident in the pattern of oxygen atoms (red spheres), which couples to the clockwise (anticlockwise) spin rotation between Mn$^{4+}$ and green (blue) Mn$^{3+}$ layers.  \textbf{(b)} A perspective view on the Mn$^{3+}$---Mn$^{4+}$---Mn$^{3+}$ configuration.  The inverse Dzyaloshinskii-Moriya interaction favours a displacement of the oxygen anions in the direction of the black arrows.  This results in an additional displacement of the Mn$^{4+}$ cation, indicated by the small yellow arrow, and thus a net polarisation along the positive $c$ axis, shown by the large yellow arrow.

\end{document}